\documentclass[twocolappendix,appendixfloats,]{emulateapj}

\usepackage[colorlinks = true, linkcolor = blue, urlcolor  = blue, citecolor = blue, anchorcolor = blue]{hyperref}
\usepackage{hyperref}
\usepackage{apjfonts}
\usepackage{rotating}
\usepackage{amsmath}
\usepackage{color}
\usepackage{graphicx}
\usepackage[dvipsnames]{xcolor}
\usepackage{CJK}

\newcommand{\te}{t_{\rm E}}
\newcommand{\thetae}{\theta_{\rm E}}
\newcommand{\pie}{\pi_{\rm E}}

\newcommand{\dl}{D_{\rm L}}
\newcommand{\ds}{D_{\rm S}}

\definecolor{darkbrown}{RGB}{139,69,19}

\shorttitle{Brown-dwarf Microlensing Events}
\shortauthors{Han et al.}

\begin{document}

\title{Candidate Brown-dwarf Microlensing Events with
Very Short Timescales and Small Angular Einstein Radii
}

\author{
Cheongho~Han\altaffilmark{0001}, 
Chung-Uk~Lee\altaffilmark{0002,101}, 
Andrzej~Udalski\altaffilmark{0003,100}, 
Andrew~Gould\altaffilmark{0004,0005,101}, 
Ian~A.~Bond\altaffilmark{0006,102},
Valerio~Bozza\altaffilmark{0035,0036}
\\
(Leading authors),\\
and \\
Michael~D.~Albrow\altaffilmark{0007}, 
Sun-Ju~Chung\altaffilmark{0002,0008},  
Kyu-Ha~Hwang\altaffilmark{0002}, 
Youn~Kil~Jung\altaffilmark{0002}, 
Yoon-Hyun~Ryu\altaffilmark{0002}, 
In-Gu~Shin\altaffilmark{0002}, 
Yossi~Shvartzvald\altaffilmark{0009}, 
Jennifer~C.~Yee\altaffilmark{0010}, 
Weicheng Zang\altaffilmark{0011},
Sang-Mok~Cha\altaffilmark{0002,0012}, 
Dong-Jin~Kim\altaffilmark{0002}, 
Hyoun-Woo~Kim\altaffilmark{0002}, 
Seung-Lee~Kim\altaffilmark{0002,0008}, 
Dong-Joo~Lee\altaffilmark{0002}, 
Yongseok~Lee\altaffilmark{0002,0012}, 
Byeong-Gon~Park\altaffilmark{0002,0008}, 
Richard~W.~Pogge\altaffilmark{0005}, 
M.~James~Jee\altaffilmark{0013,0014}, 
Doeon~Kim\altaffilmark{0001},
\\
(The KMTNet Collaboration),\\
Przemek~Mr{\'o}z\altaffilmark{0003,0031}, 
Micha{\l}~K.~Szyma{\'n}ski\altaffilmark{0003}, 
Jan~Skowron\altaffilmark{0003},
Radek~Poleski\altaffilmark{0005}, 
Igor~Soszy{\'n}ski\altaffilmark{0003}, 
Pawe{\l}~Pietrukowicz\altaffilmark{0003},
Szymon~Koz{\l}owski\altaffilmark{0003}, 
Krzysztof~Ulaczyk\altaffilmark{0015}, 
Krzysztof~A.~Rybicki\altaffilmark{0003},
Patryk~Iwanek\altaffilmark{0003}, 
Marcin~Wrona\altaffilmark{0003}\\
(The OGLE Collaboration) \\   
Fumio~Abe\altaffilmark{0016}, 
Richard~Barry\altaffilmark{0017},           
David~P.~Bennett\altaffilmark{0017,0018},
Aparna~Bhattacharya\altaffilmark{0017,0018}, 
Martin~Donachie\altaffilmark{0019}, 
Hirosane~Fujii\altaffilmark{0016},
Akihiko~Fukui\altaffilmark{0020,0021},  
Yoshitaka~Itow\altaffilmark{0016}, 
Yuki~Hirao\altaffilmark{0022}, 
Yuhei~Kamei\altaffilmark{0016},
Iona~Kondo\altaffilmark{0022}, 
Naoki~Koshimoto\altaffilmark{0023,0024}, 
Man~Cheung~Alex~Li\altaffilmark{0019},    
Yutaka~Matsubara\altaffilmark{0016}, 
Yasushi~Muraki\altaffilmark{0016}, 
Shota~Miyazaki\altaffilmark{0022}, 
Masayuki~Nagakane\altaffilmark{0022}, 
Cl\'ement~Ranc\altaffilmark{0017}, 
Nicholas~J.~Rattenbury\altaffilmark{0019}, 
Yuki~Satoh\altaffilmark{0022},
Hikaru~Shoji\altaffilmark{0022},
Haruno~Suematsu\altaffilmark{0022}, 
Denis~J.~Sullivan\altaffilmark{0025}, 
Takahiro~Sumi\altaffilmark{0022},  
Daisuke~Suzuki\altaffilmark{0026}, 
Paul~J.~Tristram\altaffilmark{0027}, 
Takeharu~Yamakawa\altaffilmark{0016},
Tsubasa~Yamawaki\altaffilmark{0022},
Atsunori~Yonehara\altaffilmark{0028}\\         
(The MOA Collaboration),\\ 
}


\affil{$^{0001}$ Department of Physics, Chungbuk National University, Cheongju 28644, Republic of Korea; cheongho@astroph.chungbuk.ac.kr} 
\affil{$^{0002}$ Korea Astronomy and Space Science Institute, Daejon 34055, Republic of Korea} 
\affil{$^{0003}$ Warsaw University Observatory, Al.~Ujazdowskie 4, 00-478 Warszawa, Poland} 
\affil{$^{0004}$ Max Planck Institute for Astronomy, K\"onigstuhl 17, D-69117 Heidelberg, Germany} 
\affil{$^{0005}$ Department of Astronomy, Ohio State University, 140 W. 18th Ave., Columbus, OH 43210, USA} 
\affil{$^{0006}$ Institute of Natural and Mathematical Sciences, Massey University, Auckland 0745, New Zealand}
\affil{$^{0007}$ University of Canterbury, Department of Physics and Astronomy, Private Bag 4800, Christchurch 8020, New Zealand} 
\affil{$^{0008}$ Korea University of Science and Technology, 217 Gajeong-ro, Yuseong-gu, Daejeon, 34113, Republic of Korea} 
\affil{$^{0009}$ Department of Particle Physics and Astrophysics, Weizmann Institute of Science, Rehovot
76100, Israel}
\affil{$^{0010}$ Center for Astrophysics $|$ Harvard \& Smithsonian 60 Garden St., Cambridge, MA 02138, USA} 
\affil{$^{0011}$ Physics Department and Tsinghua Centre for Astrophysics, Tsinghua University, Beijing 100084, China} 
\affil{$^{0012}$ School of Space Research, Kyung Hee University, Yongin, Kyeonggi 17104, Republic of Korea} 
\affil{$^{0013}$ Yonsei University, Department of Astronomy, Seoul, Republic of Korea}
\affil{$^{0014}$ Department of Physics, University of California, Davis, California, USA}
\affil{$^{0015}$ Department of Physics, University of Warwick, Gibbet Hill Road, Coventry, CV4 7AL, UK} 
\affil{$^{0031}$ Division of Physics, Mathematics, and Astronomy, California Institute of Technology, Pasadena, CA 91125, USA}
\affil{$^{0016}$ Institute for Space-Earth Environmental Research, Nagoya University, Nagoya 464-8601, Japan}
\affil{$^{0017}$ Code 667, NASA Goddard Space Flight Center, Greenbelt, MD 20771, USA}
\affil{$^{0018}$ Department of Astronomy, University of Maryland, College Park, MD 20742, USA}
\affil{$^{0019}$ Department of Physics, University of Auckland, Private Bag 92019, Auckland, New Zealand}
\affil{$^{0020}$ Instituto de Astrof\'isica de Canarias, V\'ia L\'actea s/n, E-38205 La Laguna, Tenerife, Spain}
\affil{$^{0021}$ Department of Earth and Planetary Science, Graduate School of Science, The University of Tokyo, 7-3-1 Hongo, Bunkyo-ku, Tokyo 113-0033, Japan}
\affil{$^{0022}$ Department of Earth and Space Science, Graduate School of Science, Osaka University, Toyonaka, Osaka 560-0043, Japan}
\affil{$^{0023}$ Department of Astronomy, Graduate School of Science, The University of Tokyo, 7-3-1 Hongo, Bunkyo-ku, Tokyo 113-0033, Japan}
\affil{$^{0024}$ National Astronomical Observatory of Japan, 2-21-1 Osawa, Mitaka, Tokyo 181-8588, Japan}
\affil{$^{0025}$ School of Chemical and Physical Sciences, Victoria University, Wellington, New Zealand}
\affil{$^{0026}$ Institute of Space and Astronautical Science, Japan Aerospace Exploration Agency, 3-1-1 Yoshinodai, Chuo, Sagamihara, Kanagawa, 252-5210, Japan}
\affil{$^{0027}$ University of Canterbury Mt.\ John Observatory, P.O. Box 56, Lake Tekapo 8770, New Zealand}
\affil{$^{0028}$ Department of Physics, Faculty of Science, Kyoto Sangyo University, 603-8555 Kyoto, Japan}
\affil{$^{0035}$ Dipartimento di Fisica ``E.~R.~Caianiello'', Universit\'e di Salerno, Via Giovanni Paolo II, I-84084 Fisciano (SA), Italy}
\affil{$^{0036}$ Istituto Nazionale di Fisica Nucleare, Sezione di Napoli, Via Cintia, I-80126 Napoli, Italy}
\altaffiltext{100}{OGLE Collaboration.}
\altaffiltext{101}{KMTNet Collaboration.}
\altaffiltext{102}{MOA Collaboration.}


\begin{abstract}
Short-timescale microlensing events are likely to be produced by substellar brown dwarfs 
(BDs), but it is difficult to securely identify BD lenses based on only event timescales $\te$ 
because short-timescale events can also be produced by stellar lenses with high relative 
lens-source proper motions.  In this paper, we report three strong candidate  BD-lens events 
found from the search for lensing events not only with short timescales ($\te \lesssim 6~{\rm days}$) 
but also with very small angular Einstein radii ($\thetae\lesssim 0.05~{\rm mas}$) among the 
events that have been found in the 2016--2019 observing seasons.
These events include MOA-2017-BLG-147, MOA-2017-BLG-241, and MOA-2019-BLG-256, in which the 
first two events are produced by single lenses and the last event is produced by a binary 
lens.  From the Bayesian analysis conducted with the combined $\te$ and $\thetae$ constraint, it 
is estimated that the lens masses of the individual events are 
$0.051^{+0.100}_{-0.027}~M_\odot$, 
$0.044^{+0.090}_{-0.023}~M_\odot$, and 
$0.046^{+0.067}_{-0.023}~M_\odot/0.038^{+0.056}_{-0.019}~M_\odot$ 
and the probability of the lens mass smaller than the lower limit of stars is $\sim 80\%$ 
for all events.  We point out that routine lens mass measurements of short time-scale lensing 
events require survey-mode space-based observations.
\end{abstract}

\keywords{gravitational lensing: micro -- brown dwarfs}

\section{Introduction}\label{sec:one}

Considering that brown dwarfs (BDs) share the same or similar formation mechanism as 
their heavier-mass sibling of stars and the number of stars increases as the mass 
decreases,  it may be that the Galaxy is teeming with BDs.  Due to the intrinsic 
faintness, however, it is difficult to detect BDs from imaging or spectroscopic 
observations, unless they are nearby and relatively young and/or massive. In particular, 
the Galactic bulge BD mass function cannot be probed with these techniques.  Microlensing 
provides an ideal method to detect BDs because the lensing phenomenon occurs by the 
gravity of lens objects regardless of their brightness.

In order to firmly identify BD lenses, it is required to determine lens masses.  
For general lensing events, the only observable related to the lens mass is the event 
timescale $\te$.  The event timescale is related to the physical lens parameters by
\begin{equation}
\te = {\thetae\over \mu};\ \ \
\thetae=(\kappa M \pi_{\rm rel})^{1/2};\ \ \ 
\pi_{\rm rel}={\rm au}\left({1\over \dl}-{1\over \ds} \right),
\label{eq1}
\end{equation}
where $\thetae$ is the angular Einstein radius, $\mu$ is the relative lens-source proper 
motion, $\kappa=4G/(c^2{\rm au})$, $M$ is the lens mass, and $\dl$ and $\ds$ represent the 
distances to the lens and source, respectively. Because the timescale is proportional to 
the square root of the lens mass, i.e., $\te\propto \sqrt{M}$, a considerable fraction of 
events with very short timescales are likely to be produced by BDs. However, short-timescale 
events can also be produced by stellar lenses with high relative lens-source proper motions. 
Therefore, it is difficult to firmly identify BD lenses just based on the event timescale.

For a fraction of events, it is possible to determine the angular Einstein radius, 
which is an additional observable related to the lens mass. The angular Einstein 
radius can be measured for events in which lensing lightcurves are affected by 
finite-source effects.  For events with a single lens and a single source (1L1S events), 
these effects occur when the lens passes over the surface of a source star \citep{Gould1994a}.  
See example events in \citet{Choi2012}.  For binary-lens (2L1S) events, lensing lightcurves 
are affected by finite-source effects when the source passes over the caustic. Analysis 
of the lightcurve affected by finite-source effects yields the normalized angular source 
radius $\rho$, which is related to the angular Einstein radius and angular source radius 
$\theta_*$ by $\rho=\theta_*/\thetae$. Then, the angular Einstein radius is determined 
with the additional information of the angular source radius by $\thetae=\theta_*/\rho$. 
While the event timescale is related to the three parameters of $\mu$, $\pi_{\rm rel}$, 
and $M$, the angular Einstein radius is related to only the two parameters of $\pi_{\rm rel}$ 
and $M$. Therefore, the lens mass can be better constrained with the additionally measured value 
of $\thetae$.

With the increasing observational cadence of microlensing surveys, the number of events 
with additionally measured angular Einstein radii is rapidly increasing. The duration 
of finite-source effects is approximately
\begin{equation}
\Delta t \sim {2\theta_*\over \mu}.
\label{eq2}
\end{equation}
For $\mu\sim 5~{\rm mas}~{\rm yr}^{-1}$ of typical lensing events, the duration is in 
order of hours for events associated with main-sequence source stars and $\sim 1$~day 
for events occurred on giant source stars.  With the observational cadence of 
$\sim 1$~day in the early stage of microlensing experiments, it was difficult to determine 
$\thetae$ by resolving the short-lasting parts of lensing lightcurves affected by 
finite-source effects.  With the utilization of wide-field cameras together with the 
employment of globally-distributed multiple telescopes, the observational cadence of 
lensing surveys has dramatically increased.  This enables to resolve finite-source 
lightcurves and determine angular Einstein radii for a greatly increased number of events.

In this paper, we present the analyses of three microlensing events that are very likely 
to be produced by BD lenses. For these events, the high probability of the BD lens nature 
is identified not only by the short timescales but also by the very small angular Einstein 
radii.

The paper is organized as follows.  In Section~\ref{sec:two}, we outline the procedure of 
selecting events analyzed in this work.  In Section~\ref{sec:three}, we describe the 
observations of the events and the data acquired from the observations.  We describe modeling 
the lightcurves of the individual events in Section~\ref{sec:four} and mention the 
procedure of measuring the angular Einstein radii in Section~\ref{sec:five}.  We estimate 
the masses and locations of the lenses in Section~\ref{sec:six}.  In Section~\ref{sec:seven}, 
we discuss the feasibility of measuring the microlens parallax for events similar to the 
analyzed events.  We summarize the results and conclude in Section~\ref{sec:eight}.

\begin{deluxetable*}{lccccl}
\tablecaption{Coordinates of events\label{table:one}}
\tablewidth{480pt}
\tablehead{
\multicolumn{1}{c}{Event}         &
\multicolumn{1}{c}{R.A.$_{\rm J2000}$}          &
\multicolumn{1}{c}{decl.$_{\rm J2000}$}         &
\multicolumn{1}{c}{$l$}     &
\multicolumn{1}{c}{$b$}     &
\multicolumn{1}{c}{Survey}
}
\startdata                                              
MOA-2017-BLG-147      &   17:52:09.64   &   -31:49:13.4    &  $-1^\circ\hskip-2pt.75$  &  $-2^\circ\hskip-2pt.69$   &   MOA     \\
OGLE-2017-BLG-0504    &                 &                  &     &     &   OGLE    \\
KMT-2017-BLG-0132     &                 &                  &     &     &   KMTNet  \\
\hline
MOA-2017-BLG-241      &   17:36:14.79   &   -27:02:36.0    &  $0^\circ\hskip-2pt.51$   &  $2^\circ\hskip-2pt.76$   &   MOA     \\
OGLE-2017-BLG-0776    &                 &                  &     &     &   OGLE    \\
KMT-2017-BLG-0818     &                 &                  &     &     &   KMTNet  \\
\hline
MOA-2019-BLG-256      &   18:02:11.30   &   -27:29:51.5    &  $3^\circ\hskip-2pt.09$   &  $-2^\circ\hskip-2pt.42$   &   MOA     \\
OGLE-2019-BLG-0947    &                 &                  &     &     &   OGLE    \\
KMT-2019-BLG-1241     &                 &                  &     &     &   KMTNet
\enddata                            
\tablecomments{
For a single event, there are multiple names given by the individual surveys and 
the names are listed according to the chronological order of the event discovery. 
Hereafter  we use the names given by the first discovery survey as
the representative names of the events.
\bigskip
}
\end{deluxetable*}
\bigskip


\section{Event selection}\label{sec:two}

We search for candidate BD-lens events from the sample of lensing events that have 
been found in the 2016--2019 observing seasons.  The 2016 season corresponds to the 
time of the full-scale operation of the current high-cadence lensing surveys: Optical 
Gravitational Lensing Experiment \citep[OGLE:][]{Udalski2015}, Microlensing Observations 
in Astrophysics \citep[MOA:][]{Bond2001}, and Korea Microlensing Telescope Net-work 
\citep[KMTNet:][]{Kim2016}. During this period, more than 2000 events have been 
detected each year.

Selection of candidate BD-lens events are based on the combined information of the 
event timescale and the angular Einstein radius.  For this, we first pick out 
short-timescale events, for which finite-source deviations in lensing lightcurves
are detected. In the second step, we select events with very small angular Einstein 
radii. Rough estimation of $\te$ can be easily done from the durations of events. In 
contrast, estimating $\thetae$ requires extra information of the source color, from 
which the angular source radius $\theta_*$ is estimated, and thus it is difficult to 
inspect a large sample of finite-source events.  For the efficient search for events 
with very small $\thetae$, we inspect events that are affected by severe finite-source 
effects with very large normalized source radius $\rho$.  This criterion is applied 
because the angular Einstein radius is related to the normalized source radius by 
$\thetae=\theta_*/\rho$, and thus a large $\rho$ value suggests that $\thetae$ is 
likely to be small.  We note that the shortcoming of this criterion is that it tends 
to restrict to source stars with large angular radii, i.e., giant stars, and thus limits 
the sample. For this reason, we note that there could be more events with small $\thetae$ 
from the events with lower-luminosity source stars.  In the selection of events, we impose 
requirements of $t_{\rm E}\lesssim 6$~days and $\rho\gtrsim \rho_{\rm th} \equiv 0.1$.  We 
note that the imposed threshold value $\rho_{\rm th}=0.1$ is much greater than typical values 
of events associated with main-sequence stars, $\sim (O)10^{-3}$, and giant stars, 
$\sim (O)10^{-2}$.  For events that meet these requirements, we then estimate the angular 
Einstein radii and apply another criterion of $\thetae<0.05$~mas.\footnote{For comparison, 
we note that the angular Einstein radius of a lensing event produced by a low-mass star 
with $M\sim 0.3~M_\odot$ located halfway between a source in the bulge and the observer
is about $\thetae\sim 0.5$~mas.} From this procedure, we find three candidate BD-lens events, 
including MOA-2017-BLG-147, MOA-2017-BLG-241, and MOA-2019-BLG-256, analyzed in this work.  
We note that MOA-2017-BLG-147 and MOA-2017-BLG-241 are 1L1S events and MOA-2019-BLG-256 
is a 2L1S event.

We note that there are three more events satisfying the imposed criteria besides the 
events analyzed in this work.  These events are OGLE-2016-BLG-1227, OGLE-2016-BLG-1540, 
and OGLE-2017-BLG-0560.  The lightcurve of the event OGLE-2016-BLG-1227 appears to be 
a 1L1S event affected by severe finite-source effects and the preliminary 1L1S modeling 
yields $\te\sim 3.5$~days and $\thetae\sim 0.009$~mas, making the lens a strong candidate 
of either a BD or a free-floating planet.  From detailed investigation, it is found that 
the event is produced by a wide-separation planet and the analyses will be presented in 
a separate paper.  The events OGLE-2016-BLG-1540 (with $\te\sim 0.32$~days and 
$\thetae\sim 0.009$~mas) and OGLE-2017-BLG-0560 (with $\te\sim 0.91$~days and 
$\thetae\sim 0.038$~mas) were analyzed by \citet{Mroz2018} and \citet{Mroz2019}, 
respectively.  They pointed out that the lens of OGLE-2016-BLG-1540 was likely to be a 
Neptune-mass free-floating planet in the Galactic disk and the lens of OGLE-2017-BLG-0560 
is either a Jupiter-mass free-floating planet in the disk or a BD in the bulge.

\begin{figure}
\includegraphics[width=\columnwidth]{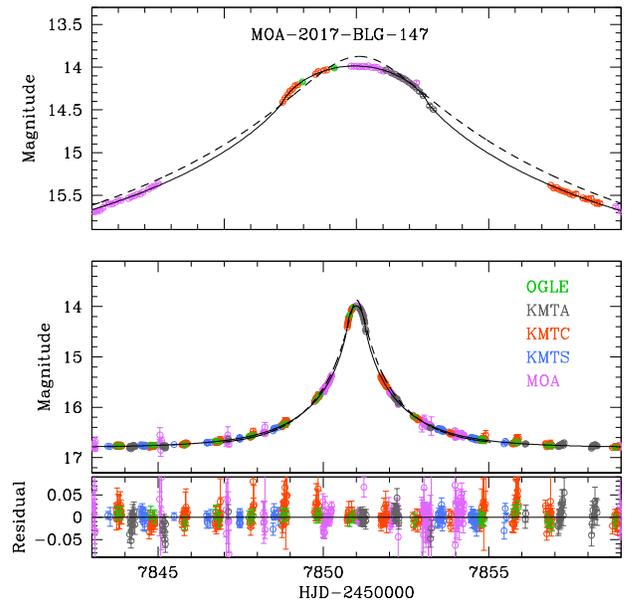}
\caption{
Lightcurve of MOA-2017-BLG-147.  
The middle panel shows the whole range of lensing 
magnification and the top panel shows the zoom of the peak region. The solid and 
dashed curves superposed on the data points represent the model curves obtained with 
and without considering finite-source effects, respectively. The colors of the data 
points are set to match those of the telescopes 
in the legend
used for the data acquisition.
The bottom panel shows the residual from the model considering finite-source effects.
\smallskip
}
\label{fig:one}
\end{figure}

\section{Observations and Data}\label{sec:three}

The analyzed lensing events share a common observational property that the lightcurves 
of the events are densely observed by the major lensing surveys despite of their short 
timescales.  All of the events are detected toward the Galactic bulge field.  In 
Table~\ref{table:one}, we list the positions of the events, both in equatorial coordinates 
(R.A., decl.)$_{\rm J2000}$ and the corresponding galactic coordinates $(l,b)$. Also 
listed are the surveys that observed the events. For each event, different names are 
given by the individual surveys, and we list all the names according to the chronological 
order of the event discovery. Hereafter, we use the names given by the first discovery 
survey as the representative names of the events.

\begin{figure}
\includegraphics[width=\columnwidth]{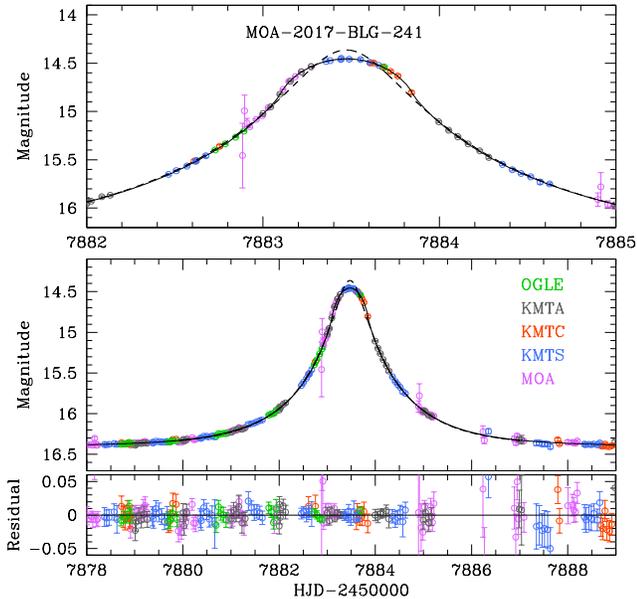}
\caption{
Lightcurve of MOA-2017-BLG-241.  Notations are same as those in Fig.~\ref{fig:one}.
\bigskip
}
\label{fig:two}
\end{figure}

The survey observations were conducted using multiple telescopes that were equipped 
with wide-field cameras and globally distributed in the southern hemisphere. The 
telescope used for the OGLE survey is located at the Las Campanas Observatory in Chile. 
The telescope has a 1.3~m aperture, and it is equipped with a mosaic camera that consists 
of 32 chips with each chip composed of $2{\rm k}\times 4{\rm k}$ pixels. The camera covers 
a $1.4~{\rm deg}^2$ field of view with a single exposure.  The MOA 1.8~m telescope, located 
at the Mt.~John Observatory in New Zealand, is equipped with a camera that consists of ten 
$2{\rm k}\times2{\rm k}$ chips with a total $2.2~{\rm deg}^2$ field of view. The KMTNet 
observations were carried out using three identical 1.6~m telescopes located at the Siding 
Spring Observatory in Australia (KMTA), Cerro Tololo Interamerican Observatory in Chile 
(KMTC), and the South African Astronomical Observatory in South Africa (KMTS).  The camera 
mounted on each of the KMTNet telescopes consists of four $9{\rm k}\times 9{\rm k}$ chips 
with a total $4~{\rm deg}^2$ field of view. The wide field of view of the surveys using the 
globally distributed telescopes enable dense and continuous coverage of the events 
despite their short durations. Observations by the OGLE and KMTNet surveys were conducted 
mostly in $I$ band with occasional observations in $V$ band. MOA observations were carried 
out in a customized broad $R/I$ filter.

Reduction of the data sets is conducted using the photometry codes developed by the 
individual survey groups based on the difference imaging method \citep{Alard1998}: 
\citet{Wozniak2000} (OGLE), \citet{Bond2001} (MOA), and \citet{Albrow2009} (KMTNet). 
For a subset of the KMTC data set, additional photometry is conducted using the pyDIA 
code \citep{Albrow2017} for the source color measurement. We readjust the error bars 
of the individual data sets following the method described in \citet{Yee2012}.

\begin{figure}
\includegraphics[width=\columnwidth]{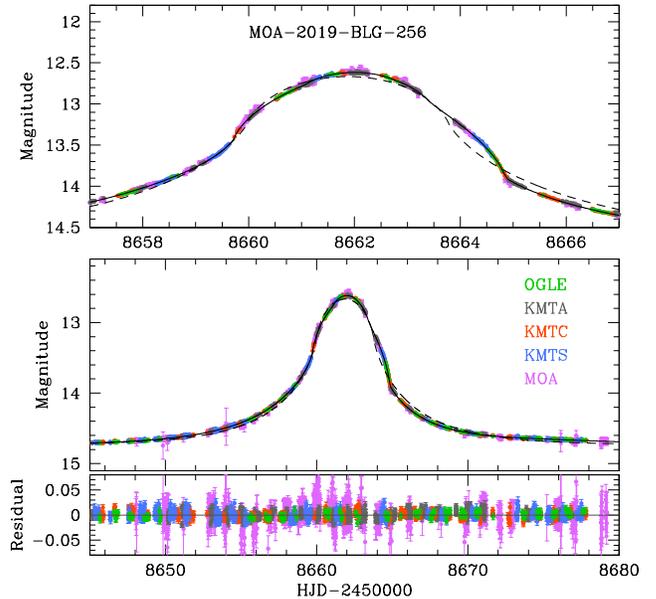}
\caption{
Lightcurve of MOA-2019-BLG-256.  The solid and dashed curves represent the
model curves based on 2L1S and 1L1S
modeling, respectively. For both models, finite-source
effects are considered.
\smallskip
}
\label{fig:three}
\end{figure}

In Figures~\ref{fig:one} through \ref{fig:three}, we present the lightcurves of the 
MOA-2017-BLG-147, MOA-2017-BLG-241, and MOA-2019-BLG-256, respectively. 
As mentioned, the lightcurves of all events are affected by severe finite-source effects, 
and the peak regions show strong deviations from the point-source lightcurves (dashed curves). 
To better show the lightcurve deviation affected by finite-source effects, we present the 
zoom of the peak region in the upper panel of each figure. At first glance, the lightcurve 
of MOA-2019-BLG-256 appears to be similar to those of the other events produced by 
finite-source 1L1S events, but a close look shows asymmetry with respect to the peak. As we 
will show in the following section, the event is produced by a binary lens.

\begin{deluxetable*}{lccc}
\tablecaption{Best-fit lensing parameters\label{table:two}}
\tablewidth{480pt}
\tablehead{
\multicolumn{1}{c}{Parameter}            &
\multicolumn{1}{c}{MOA-2017-BLG-147}     &
\multicolumn{1}{c}{MOA-2017-BLG-241}     &
\multicolumn{1}{c}{MOA-2019-BLG-256}     
}
\startdata                                              
$t_0$ (${\rm HJD}^\prime$) &   7850.994 $\pm$ 0.001   &  7883.473 $\pm$ 0.001   &  8662.089 $\pm$ 0.001  \\
$u_0$                      &   0.092 $\pm$ 0.001      &  0.211 $\pm$ 0.005      &  0.076 $\pm$ 0.001     \\
$t_{\rm E}$   (days)       &   2.679 $\pm$ 0.023      &  1.868 $\pm$ 0.023      &  8.723 $\pm$ 0.008     \\
$t_{\rm E,1}$ (days)       &   -                      &  -                      &  6.439 $\pm$ 0.006     \\
$t_{\rm E,2}$ (days)       &   -                      &  -                      &  5.884 $\pm$ 0.005     \\
$s$                        &   -                      &  -                      &  1.968 $\pm$ 0.002     \\
$q$                        &   -                      &  -                      &  0.835 $\pm$ 0.003     \\
$\alpha$ (rad)             &   -                      &  -                      &  2.313 $\pm$ 0.001     \\
$\rho$                     &   0.137 $\pm$ 0.001      &  0.290 $\pm$ 0.005      &  0.213 $\pm$ 0.001     \\
$f_{\rm s,OGLE}$           &   3.076                  &  4.345                  &  21.310                \\   
$f_{\rm b,OGLE}$           &   -0.062                 &  0.010                  &  -1.296                
\enddata                            
\tablecomments{${\rm HJD}^\prime={\rm HJD-2450000}$.
For the 2L1S event MOA-2019-BLG-256,
$\te$ is the event timescale corresponding to the total mass of the binary lens, and
$t_{\rm E,1}$ and $t_{\rm E,2}$ represent the timescales corresponding to the
masses of individual lens components.
\smallskip
}
\end{deluxetable*}

\section{Modeling lightcurves}\label{sec:four}

The first step for the analyses of the events is conducting modeling on the observed 
lightcurves.  Lightcurve modeling is carried out by searching for a set of the lensing parameters 
that best describes the observed lightcurves. For a 1L1S event with a point source, the 
lensing lightcurve is described by three parameters of $t_0$, $u_0$, and $\te$ \citep{Paczynski1986}.
The first two of these parameters represent the time of the closest lens-source approach 
and the lens-source separation (normalized to $\thetae$) at that time, i.e., impact parameter, 
respectively.  For a 1L1S event in which the source radius is greater than the impact parameter, 
i.e., $\rho> u_0$, the lensing lightcurve is affected by finite-source effects. For the 
description of such events, one needs an additional lensing parameter of $\rho$. For 2L1S 
events, one needs additional parameters to describe the binary nature of the lens. These 
additional parameters include $s$, $q$, and $\alpha$. The parameter $s$ denotes the projected 
separation (normalized to $\thetae$), $q$ represents the mass ratio between the binary lens 
components, and $\alpha$ represents the incidence angle of the source trajectory with respect 
to the binary axis.

Lensing magnifications affected by finite-source effects differ from those of a point 
source. For 1L1S events, we compute finite-source magnifications using the semianalytic 
expressions first derived by \citet{Gould1994a} and \citet{Witt1994} and later refined by 
\citet{Yoo2004}.  These approximation may not be valid in the region of a very large $\rho$, 
and thus we check the validity of the expressions by comparing magnifications computed by 
using a contouring method.  We find that the semianalytic expressions are valid in the cases 
of the analyzed events.  For 2L1S events, we compute magnifications using the numerical 
ray-shooting method described in \citet{Dong2006}.  In computing finite-source magnifications, 
we consider the variation of the source surface brightness caused by limb darkening. To account 
for the limb-darkening variation, we model the surface brightness of the source star as 
\begin{equation} 
S_\lambda = \bar{S}_{\lambda} \left[ 1- \Gamma_\lambda 
\left( 1-{3\over 2}\cos\theta \right) \right].  \label{eq3} 
\end{equation} 
Here $\bar{S}_\lambda$ denotes the mean surface brightness, $\Gamma_\lambda$ is the linear 
limb-darkening coefficient, and $\theta$ represents the angle between the line of sight 
toward the source center and the normal to the source surface. The limb-darkening 
coefficients are estimated based on the stellar types of the source stars. As we will 
show in the following section, the source stars of the analyzed events are giant stars 
of a similar spectral type ranging from K0 to K3. Based on the stellar type, we set the 
limb-darkening coefficients as $\Gamma_I=0.41$, and 
$\Gamma_{\rm MOA}\sim (\Gamma_I+\Gamma_R)/2=0.52$ by adopting the values from \citet{Claret2000} 
under the assumption that $v_{\rm turb}=2~{\rm km}~{\rm s}^{-1}$, $\log(g/g_\odot)=-2.4$, 
and $T_{\rm eff}=4500~K$.

\begin{figure}
\includegraphics[width=\columnwidth]{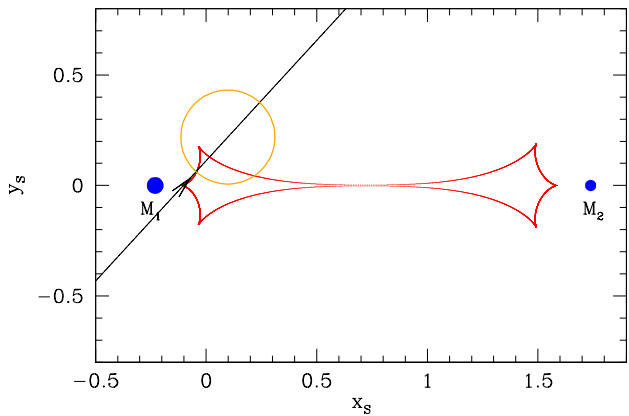}
\caption{
Lens system configuration of the 2L1S event MOA-2019-BLG-256.
The two blue dots, marked by $M_1$ and $M_2$, represent the positions of the lens
components and the cuspy closed figure is the caustic. The line with an arrow
is the source trajectory and the orange circle on the trajectory represents the
relative size of the source. All lengths are scaled to the angular Einstein radius
corresponding to the total mass of the binary lens.
\smallskip
}
\label{fig:four}
\end{figure}

We search for the best-fit lensing parameters using the combination of downhill and 
grid-search approaches. For events produced by single lenses, i.e., 
MOA-2017-BLG-147, and MOA-2017-BLG-241, lensing parameters are searched for using the 
downhill approach based on the Markov Chain Monte Carlo (MCMC) method. In this search, 
the initial values of the parameters are given considering the time of the peak, $t_0$, 
peak magnification, $A_{\rm peak}$, duration of the event, and duration of finite-source 
anomaly, $\Delta t$. For 1L1S events affected by severe finite-source effects, the peak 
magnification is approximated as
$A_{\rm peak}\sim (1+4/\rho^2)^{1/2}$ \citep{Maeder1973, Riffeser2006, 
Agol2003, Han2016}.  
For the 2L1S event, i.e., MOA-2019-BLG-256, the analysis is done 
in two steps. In the first step, we conduct grid search for the binary lensing parameters 
$s$ and $q$, while the other parameters are searched for using the MCMC downhill approach. 
In the second step, we refine the solution(s) found from the initial grid search by 
allowing all parameters including $s$ and $q$ to vary.  Modeling 2L1S events often 
results in multiple solutions caused by various types of degeneracy. For MOA-2019-BLG-256, 
we find a unique solution without any degeneracy.
We also check the possible degeneracy between binary-lens (2L1S) and binary-source (1L2S) 
solutions. We find that the 1L2S interpretation does not explain the observed anomaly.

In Table~\ref{table:two}, we list the best-fit lensing parameters of the individual events. 
For the 2L1S event MOA-2019-BLG-256, we present three event timescales of ($t_{\rm E}$, 
$t_{\rm E,1}$, $t_{\rm E,2}$), in which $\te$ is the timescale corresponding to the total 
mass of the binary lens, while $t_{\rm E,1}$ and $t_{\rm E,2}$ represent the timescales 
corresponding to the masses of individual lens components, i.e., $t_{\rm E,1}=\sqrt{1/(1+q)}\te$ 
and $t_{\rm E,2}=\sqrt{q/(1+q)}\te$. The uncertainties of the parameters are estimated as the 
standard deviation of the points in the MCMC chain. It is found that the estimated event 
timescales are very short, ranging from $\te\sim 1.9$~days to $\sim 6.4$~days according to 
the timescales corresponding to the individual lens components.  It is also found that the 
normalized source radii are very big, ranging from $\rho\sim 0.14$ to $\sim 0.29$.  Also 
listed in the table are the flux values of the source, $f_{\rm s,OGLE}$, and blend, 
$f_{\rm b,OGLE}$, estimated according to the OGLE scale, in which $f=1$ for an $I=18.0$ mag 
star.  The dominance of the source flux over the blend flux indicates that blending is 
negligible for all events.

\begin{deluxetable*}{lccc}
\tablecaption{Best-fit lensing parameters\label{table:three}}
\tablewidth{480pt}
\tablehead{
\multicolumn{1}{c}{Parameter}            &
\multicolumn{1}{c}{MOA-2017-BLG-147}     &
\multicolumn{1}{c}{MOA-2017-BLG-241}     &
\multicolumn{1}{c}{MOA-2019-BLG-256}     
}
\startdata                                              
$V-I$                    &    $2.93\pm 0.07$      &  $2.84\pm 0.03$     &  $2.48\pm 0.01$   \\
$I$                      &    $16.59\pm 0.01$     &  $16.72\pm 0.01$    &  $15.32\pm 0.01$  \\
$(V-I,I)_{\rm RGC}$      &    $(3.00,17.03)$      &  $(2.61,17.15)$     &  $(2.30,16.67)$   \\
$(V-I,I)_{\rm RGC,0}$    &    $(1.06,14.51)$      &  $(1.06,14.65)$     &  $(1.06,14.30)$   \\
$(V-I)_0$                &    $0.99\pm 0.07$      &  $1.30\pm 0.03$     &  $1.25\pm 0.01$   \\
$I_0$                    &    $14.03\pm 0.01$     &  $14.22\pm 0.01$    &  $12.95\pm 0.10$  \\
$\theta_*$ ($\mu$as)     &    $6.94\pm 0.69$      &  $8.06\pm 0.60$     &  $14.07\pm 0.99$  \\
$\theta_{\rm E}$ (mas)   &    $0.051\pm 0.005$    &  $0.028\pm 0.004$   &  $0.066\pm 0.005$ \\
$\theta_{\rm E,1}$ (mas) &    -                   &  -                  &  $0.049\pm 0.004$ \\
$\theta_{\rm E,2}$ (mas) &    -                   &  -                  &  $0.045\pm 0.003$ \\
$\mu$ (mas yr$^{-1}$)    &    $6.89\pm 0.69$      &  $5.42\pm 0.83$     &  $2.76\pm 0.19$   \\
Spectral type            &    K0III               &  K3III              &  K3III
\enddata                            
\tablecomments{
For the 2L1S event MOA-2019-BLG-256,
$\thetae$ is the angular Einstein radius corresponding to the total mass of the binary lens, and
$\theta_{\rm E,1}$ and $\theta_{\rm E,2}$ represent the Einstein radii corresponding to the
masses of individual lens components.
\bigskip
  }
\end{deluxetable*}

In Figure~\ref{fig:four}, we present lens system configuration of the 2L1S event MOA-2019-BLG-256.  
The blue dot marked by $M_1$ and $M_2$ denote the positions of the binary lens components.  
The mass ratio between the lens components is $q=M_2/M_1=0.835 \pm 0.003$, and they are separated 
in projection by $s=1.968 \pm 0.002$.  The cuspy curves represent the caustic. Because the 
separation between $M_1$ and $M_2$ is greater than $\thetae$, i.e., $s>1.0$, the caustic is 
composed of two segments, which are located close to the individual lens components. The line 
with an arrow represents the source trajectory. The orange circle on the source trajectory is 
marked to represent the source size with respect to the caustic. It is found that the size of 
the source is similar to that of the caustic located close to $M_1$. The source approaches and 
crosses the caustic multiple times. For general events with a source much smaller than a caustic, 
sharp spike features appear in the lensing lightcurve at the times of the individual caustic 
approaches and crossings.  For MOA-2019-BLG-256, such a spike feature does not appear in the 
light curve due to the severe attenuation of the lensing magnification by finite-source effects.

\section{Angular Einstein Radius}\label{sec:five}

For the additional constraint of the lens mass, we estimate the angular Einstein radii of the 
events.  The angular Einstein radius is estimated from the combination of the normalized source 
radius $\rho$ and the angular source radius $\theta_*$ by $\thetae=\theta_*/\rho$.  The value of 
$\rho$ is measured from modeling the parts of the lightcurve affected by finite-source effects. 
The angular source radius is estimated from the de-reddened color $(V-I)_0$ and brightness $I_0$ 
of the source star using the method of \citet{Yoo2004}. Following this method, we first measure 
the instrumental color $V-I$ and magnitude $I$ of the source and place the source location on the 
instrumental color-magnitude diagram (CMD). We then measure the offsets in color, $\Delta (V-I)$, 
and magnitude, $\Delta I$, from the centroid of red giant clump (RGC) with a location on the CMD 
of $(V-I,I)_{\rm RGC}$. We then estimate the calibrated de-reddened source color and magnitude, 
$(V-I,I)_0$, using the known de-reddened values of the RGC centroid by
\begin{equation}
(V-I,I)_0 = (V-I,I)_{\rm RGC,0}+\Delta(V-I,I).
\label{eq4}
\end{equation}
Here $(V-I,I)_{\rm RGC,0}$ represent the de-reddened color
and magnitude of the RGC centroid \citep{Bensby2013, Nataf2013}.

\begin{figure}
\includegraphics[width=\columnwidth]{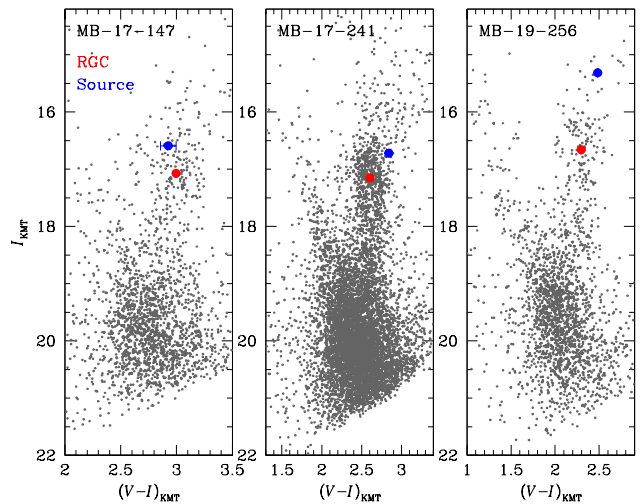}
\caption{
Source locations (blue dots) with respect to the centroids of red giant clump (RGC, red 
dots) in the instrumental color-magnitude diagrams constructed based on the pyDIA photometry 
of the KMTC data set.
\bigskip
}
\label{fig:five}
\end{figure}

\begin{figure*}
\epsscale{0.95}
\plotone{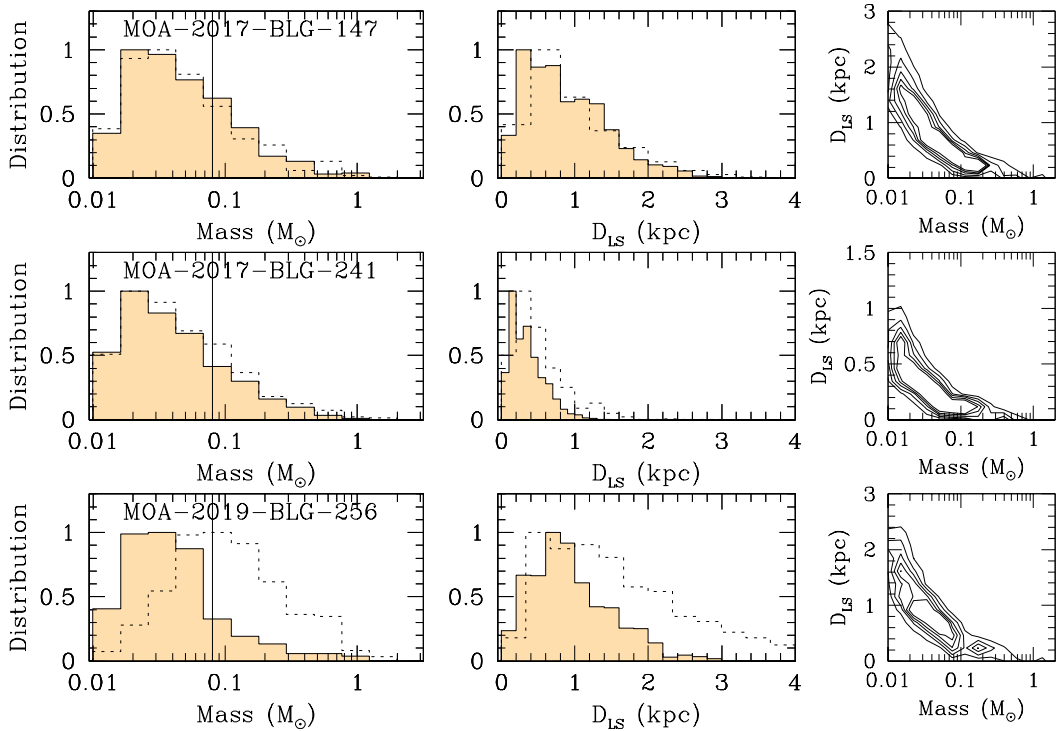}
\caption{
Probability distributions of the lens mass ($M$) and the lens-source separation ($D_{\rm LS}$) 
obtained from Bayesian analyses.  In each panel, the solid curve is the probability distribution 
obtained with the combined $\te+\thetae$ constraint, whereas the dotted curve is obtained using 
the constraint of only $\te$.  The vertical line in each left panel indicates the boundary between 
stars and BDs, i.e., $0.08~M_\odot$. For the 2L1S event MOA-2019-BLG-256, the mass distribution 
is for the heavier lens component, $M_1$.
\smallskip
}
\label{fig:six}
\end{figure*}

In Figure~\ref{fig:five}, we mark the positions of the source stars of the individual events 
with respect to the RGC centroids on the instrumental CMDs. The CMDs are constructed based on 
the pyDIA photometry of the KMTC data. 
In Table~\ref{table:three}, we 
list the colors and magnitudes of the source, $(V-I,I)$, and the RGC centroid, $(V-I,I)_{\rm RGC}$, 
on the instrumental CMD.  Also listed are the de-reddened colors and magnitudes of the RGC centroid, 
$(V-I,I)_{\rm RGC,0}$, toward the fields of the individual events \citep{Nataf2013}.  We note that 
$I_{\rm RGC,0}$ slightly varies because the distance to the RGC centroid varies depending on the 
galactic longitude $l$ due to the tilt of the triaxial bulge with respect to the line of sight.  
With $(V-I,I)_{\rm RGC,0}$ together with the measured offsets $\Delta(V-I,I)$, the de-reddened 
colors and magnitudes of the source stars are computed using equation~(\ref{eq4}) and listed in 
Table~\ref{table:three}. The ranges of the $I$-band magnitudes, $13.0\lesssim I_0\lesssim 14.2$, 
and the color, $1.0\lesssim (V-I)_0\lesssim 1.3$, indicate that the source stars of the events 
are bulge giant stars of a similar spectral type, ranging from K0 to K3.

With the estimated de-reddened color and magnitude, we then determine the angular
source radii. This is done first by converting the measured $V-I$ color into $V-K$ color using
the color-color relation of \citet{Bessell1988} and then estimating $\theta_*$ using the 
$(V-K)/\theta_*$ relation of \citet{Kervella2004}. Once the source radius is estimated, the angular 
Einstein radius is determined by $\thetae=\theta_*/\rho$.

In Table~\ref{table:three}, we list the estimated values of $\theta_*$ and $\thetae$ for the 
individual events.  For the 2L1S event MOA-2019-BLG-256, we additionally present the Einstein 
radii corresponding to the masses of the individual lens components, $\theta_{\rm E,1}$, and 
$\theta_{\rm E,2}$, similar to the presentation of $t_{\rm E,1}$ and $t_{\rm E,2}$ in 
Table~\ref{table:two}.  Also listed are the relative lens-source proper motions estimated by
\begin{equation}
\mu = {\thetae \over \te}.
\label{eq5}
\end{equation}
It is found that the angular Einstein radii are in the range of 
$0.028~{\rm mas}\lesssim \thetae \lesssim 0.051~{\rm mas}$. These values are more than an order 
smaller than $\sim 0.5$~mas of typical lensing events produced by low-mass lenses located roughly 
halfway between the observer and source.  The estimated relative lens-source proper motions are 
in the range of $2.8~{\rm mas}~{\rm yr}^{-1}\lesssim \mu \lesssim 6.9~{\rm mas}~{\rm yr}^{-1}$.  
These values are smaller or similar to $\sim 5~{\rm mas}~{\rm yr}^{-1}$ of typical lensing events.  
This indicates that the very short timescales of the analyzed events are not caused by unusually 
high relative lens-source proper motions, but more likely to be caused by the low masses of the 
lenses.

\section{Nature of Lenses}\label{sec:six}

For the characterization of the lenses, we estimate the physical lens parameters of the lens mass
$M$ and distance $\dl$. In order to uniquely determine $M$ and $\dl$, it is required to determine 
both the angular Einstein radius $\thetae$ and the microlens parallax $\pie$, which are related to 
the lens mass and distance by
\begin{equation}
M={\thetae\over \kappa\pie};\qquad \dl = {{\rm au}\over \pie\thetae+ \pi_{\rm S}},
\label{eq6}
\end{equation}
where $\pi_{\rm S}={\rm au}/D_{\rm S}$ is the parallax of the source.  For all the analyzed events, 
the angular Einstein radii are securely measured from the detections of finite-source effects. The 
microlens parallax is measurable by detecting deformations in lensing lightcurves caused by the 
deviation of the source motion from rectilinear due to the change of the observer's position induced 
by the orbital motion of Earth around the sun \citep{Gould1992}, e.g., OGLE-2016-BLG-0156 \citep{Jung2019}.
For none of the events, the microlens 
parallax can be measured through this annual microlens-parallax channel because the timescales of the 
events are too short to yield measurable deviations in the lensing lightcurves. Besides this channel, 
the microlens parallax can be measured from simultaneous observations of lensing events using 
ground-based telescopes and a space-based satellite: ``space-based microlens parallax'' 
\citep{Refsdal1966, Gould1994b}, e.g., OGLE-2015-BLG-0966 \citep{Street2016}.
See more detailed discussion about the space-based microlens-parallax 
measurements in section~\ref{sec:seven}.  Unfortunately, space-based observation has been conducted for 
none of the events. We, therefore, estimate the physical lens parameters by conducting Bayesian analysis 
of the events with the constraints of the measured event timescales together with the angular Einstein 
radii.

\begin{deluxetable}{lcc}
\tablecaption{Source proper motion\label{table:four}}
\tablewidth{240pt}
\tablehead{
\multicolumn{1}{c}{Event}                                     &
\multicolumn{1}{c}{$\mu_{\rm R.A.}$  (mas~yr$^{-1}$)}         &
\multicolumn{1}{c}{$\mu_{\rm decl.}$ (mas~yr$^{-1}$)}     
}
\startdata                                              
MOA-2017-BLG-147  &  $-5.348 \pm 0.335$   &  $-7.694 \pm  0.272$ \\
MOA-2017-BLG-241  &  $-3.775 \pm 0.450$   &  $-4.049 \pm  0.396$ \\
MOA-2019-BLG-256  &  $-2.299 \pm 0.170$   &  $-6.973 \pm  0.134$  
\enddata                            
\tablecomments{
$\mu_{\rm R.A.}$ and $\mu_{\rm decl.}$ denote the proper motions in 
right ascension and declination directions, respectively.
}
\end{deluxetable}

Bayesian analyses are conducted based on the prior models of the physical and dynamical 
distributions of astronomical objects in the Galaxy and their mass function.  For the 
physical distributions, we use the model described in sections 2.1 and 2.2 of \citet{Han2003}.  
For the model of the relative lens-source motion, we adopt the non-rotating barred bulge model 
described in table~1 of \citet{Han1995}. For the mass function of lenses, we use the 
\citet{Chabrier2003} model for stars and BDs and the \citet{Gould2000} model for stellar 
remnants, i.e., black holes, neutron stars, and white dwarfs.  With these models, we conduct 
Monte Carlo simulation to produce numerous ($4\times 10^{7}$) artificial lensing events, from 
which the probability distributions of the lens mass and distance are obtained. We obtain two 
sets of probability distributions, in which one set of distributions are obtained with only 
the constraint of $\te$, whereas the other set of distributions are obtained with the combined 
$\te$ and $\thetae$ constraint.  We note that the source stars of all the events are bright 
and their proper motions are measured by {\it Gaia} \citep{Gaia2018}. In Table~\ref{table:four}, 
we list the proper motions of the individual events.  We consider the measured proper motions 
of the source stars in the Bayesian analysis.

In Figure~\ref{fig:six}, we present the probability distributions of the physical lensing 
parameters obtained from the Bayesian analysis.  For each event, the left and middle panels 
show the distributions of the lens mass and the lens-source separation ($D_{\rm LS}$), 
respectively, and the right panel show the contours of the probability on the $M$--$D_{\rm LS}$ 
plane.  The contours are drawn at the levels of 0.1, 0.2, 0.3, 0.4, and 0.5 with respect to the 
maximum probability.  We note that the lenses are located very close to the source in all 
cases of the events and thus we present the distribution of $D_{\rm LS}$ instead of $\dl$. 
The solid and dotted curves represent the distributions obtained with $\te+\thetae$ and $\te$ 
constraints, respectively. In Table~\ref{table:five}, we list the estimated physical lens 
parameters. For the 2L1S event MOA-2019-BLG-256, we list the masses of both lens components, 
i.e., $M_1$ and $M_2$.  The presented value of each parameter is estimated as the median of 
the probability distribution and the lower and upper uncertainties are estimated as the 16\% 
and 84\% of the distribution, respectively.

\begin{deluxetable}{lccc}
\tablecaption{Physical lens parameters\label{table:five}}
\tablewidth{240pt}
\tablehead{
\multicolumn{1}{c}{Event}                        &
\multicolumn{1}{c}{$M_1$ ($M_\odot$)}            &
\multicolumn{1}{c}{$M_2$ ($M_\odot$)}            &
\multicolumn{1}{c}{$D_{\rm LS}$ (kpc)}     
}
\startdata                                              
MOA-2017-BLG-147  &  $0.051^{+0.100}_{-0.027}$  &  -                          &  $0.87^{+0.67}_{-0.45}$ \\
MOA-2017-BLG-241  &  $0.044^{+0.090}_{-0.023}$  &  -                          &  $0.36^{+0.28}_{-0.18}$ \\
MOA-2019-BLG-256  &  $0.046^{+0.067}_{-0.023}$  &  $0.038^{+0.056}_{-0.019}$  &  $0.94^{+0.62}_{-0.46}$  
\enddata                            
\tablecomments{For the 2L1S event MOA-2019-BLG-256, $M_1$ and $M_2$ denote the masses of
the individual lens components.
\smallskip
}
\end{deluxetable}

We find that the lenses of all events share similar properties that they are very likely to 
be substellar objects located very close to the source stars.  From the Bayesian analysis, 
it is estimated that the masses of the lenses are 
$0.051^{+0.100}_{-0.027}~M_\odot$, 
$0.044^{+0.090}_{-0.023}~M_\odot$, and 
$0.046^{+0.067}_{-0.023}~M_\odot/0.038^{+0.056}_{-0.019}~M_\odot$
for MOA-2017-BLG-147L, MOA-2017-BLG-241L, and MOA-2019-BLG-256LAB, respectively.  The 
probability for the lens mass smaller than the lower limit for the mass of a star is 
about 80\% for all events.  The lenses of the individual events are located at the 
locations with the distances from the source of 
$D_{\rm LS}=
 0.87^{+0.67}_{-0.45}$~kpc, 
$0.36^{+0.28}_{-0.18}$~kpc, and 
$0.94^{+0.62}_{-0.46}$~kpc.  
The estimated lens masses and locations indicate that the lenses of the events are bulge 
BDs located close to the source stars.  We note that MOA-2019-BLG-256LAB is the fourth 
microlensing BD binary followed by OGLE-2009-BLG-151L, OGLE-2011-BLG- 0420L \citep{Choi2013}, 
and OGLE-2016-BLG-1469L \citep{Han2017}.

It is found that the additional constraint provided by the angular Einstein radius helps 
to reveal the substellar nature of the lenses.  For MOA-2017-BLG-147 and MOA-2017-BLG-241, 
the probability distributions of $M$ and $D_{\rm LS}$ with the additional constraint of 
$\thetae$ are not much different from the distributions obtained with only $\te$ constraint, 
indicating that the additional constraint of $\thetae$ is not very strong.  However, for 
MOA-2019-BLG-256, the additional constraint of $\thetae$ substantially shifts the most 
probable lens mass and location toward lower masses and closer to the source, respectively.  
For the former two events, the event timescales, $\te<2.7$~days, are very short and thus the 
timescale alone constrains that the lens is likely to be a substellar object.  On the other 
hand, the event timescales of MOA-2019-BLG-256, $\te\sim 8.7$~days, is relatively long and 
the BD nature of the lens can be constrained with the additional constraint of the very small 
$\thetae$.  The very small $\thetae$ values also tightly constraint the lens locations, i.e., 
very close to the source, because $\thetae\propto (D_{\rm LS}/D_{\rm L}D_{\rm S})^{1/2}$.

\section{Discussion}\label{sec:seven}

Although the probability of the lenses to be BDs is high, the ranges of the lens masses 
estimated from the Bayesian analysis are rather big. To firmly identify the BD nature of 
the lenses, it is desirable to uniquely determine the lens masses by additionally measuring 
the values of the microlens parallax.

We point out that the microlens parallax values and thus the lens masses of the events 
could have been uniquely determined if the events had been observed using a satellite 
separated from Earth by a substantial fraction of an au.  Space-based microlens-parallax 
measurement is optimized when the projected Earth-satellite separation as seen from the 
lens-source line of sight (projected satellite separation), $D_\perp$, comprises an important 
portion of the physical Einstein radius projected onto the plane of the observer (projected 
Einstein radius), $\tilde{r}_{\rm E}=(D_{\rm S}/D_{\rm LS})r_{\rm E}$.  Here 
$r_{\rm E}=D_{\rm L}\thetae$ represents the physical Einstein radius.  If 
$D_\perp \gg \tilde{r}_{\rm E}$, the lensing magnifications observed by ground-based 
telescopes would be difficult to be observed by a space-based satellite because the impact 
parameter of the lens-source approach seen from the satellite would be too big to induce 
lensing magnifications.  If $D_\perp \ll \tilde{r}_{\rm E}$, in contrast, the difference 
between the two lensing lightcurves obtained from the ground- and space-based observations 
would be too small to securely measure $\pie$.

Considering the {\it Spitzer} telescope as an example of a satellite in a heliocentric orbit, 
we estimate the values of $r_{\rm E}$, $\tilde{r}_{\rm E}$, and $D_\perp$ and list them in 
Table~\ref{table:six}.  We note that the projected Einstein radius $\tilde{r}_{\rm E}$ is much 
bigger than $r_{\rm E}$ because $\tilde{r}_{\rm E}$ is inversely proportional to the lens-source 
distance, i.e., $\tilde{r}_{\rm E}=(D_{\rm S}/D_{\rm LS})r_{\rm E}$, and the lens-source separations 
are very small for the analyzed events.  We also list the ratios of $D_\perp/\tilde{r}_{\rm E}$ 
corresponding to the {\it Spitzer} telescope locations at the times of the events.  The ratios 
are in the range of $0.3 \lesssim D_\perp/\tilde{r}_{\rm E} \lesssim 0.6$, which are optimal 
ratios for secure $\pie$ measurements.

\begin{deluxetable}{lcccc}
\tablecaption{Projected Einstein radius\label{table:six}}
\tablewidth{240pt}
\tablehead{
\multicolumn{1}{c}{Event}                               &
\multicolumn{1}{c}{$r_{\rm E}$ (au)}                    &
\multicolumn{1}{c}{$\tilde{r}_{\rm E}$ (au)}            &
\multicolumn{1}{c}{$D_\perp$ (au)}                      &
\multicolumn{1}{c}{$\tilde{r}_{\rm E}/D_\perp$ (au)}     
}
\startdata                                              
MOA-2017-BLG-147  & 0.36   & 3.3   & 1.59    & 0.48     \\
MOA-2017-BLG-241  & 0.21   & 4.8   & 1.59    & 0.33     \\
MOA-2019-BLG-256  & 0.35   & 2.9   & 1.73    & 0.60      
\enddata                            
\tablecomments{
$\tilde{r}_{\rm E}$ denotes the physical Einstein radius projected onto the plane of the observer
and $D_\perp$ represents the projected Earth-{\it Spitzer}
separation as seen from the lens-source line of sight.
\smallskip
  }
\end{deluxetable}

For none of the events, {\it Spitzer} observation could be conducted because of the 
combined reasons that the current {\it Spitzer} microlensing campaign \citep{Calchi2015} 
has been conducted in a follow-up mode together with the fact that the timescales of the 
events are very short. According to the protocol of the {\it Spitzer} sample selection 
\citep{Yee2015}, very short-timescale events are unlikely to be selected because immediate 
follow-up observation is difficult due to the relatively long period (a week) of uploading 
observation sequences and the time required to prepare the sequences. These difficulties 
of observing short-timescale events can be overcome if space-based observations are carried 
in a survey mode simultaneously with a ground-based survey.  Another important reason for 
difficulty of observing the events is the short time window, $\sim 40$~days, through which 
the bulge field is observable simultaneously from {\it Spitzer} and from the ground.  The 
{\it Spitzer} window ran during 7927--7969 and 8671--8712 in the 2017 and 2019 seasons, 
respectively.  As a result, all of the events were at (or nearly at) baseline by the time 
{\it Spitzer} observations started.

\section{Summary and conclusion}\label{sec:eight}

We investigated strong candidate BD-lens events found from the search for lensing events 
not only with short timescales but also with very small angular Einstein radii.  By imposing 
the criteria of $\te \lesssim 6~{\rm days}$ and $\thetae\lesssim 0.05~{\rm mas}$ for events 
detected since the 2016 season, we found three events including  MOA-2017-BLG-147, MOA-2017-BLG-241, 
and MOA-2019-BLG-256, in which the lens of the last event is a binary.  By measuring the event 
timescales and angular Einstein radii from  lightcurve modeling followed by Bayesian analyses 
of the events with the combined constraint of $\te$ and $\thetae$, we estimated that the lens 
masses of the individual events were
$0.051^{+0.100}_{-0.027}~M_\odot$, 
$0.044^{+0.090}_{-0.023}~M_\odot$, and 
$0.046^{+0.067}_{-0.023}~M_\odot/0.038^{+0.056}_{-0.019}~M_\odot$.
We pointed out that uniquely determining lens masses of short timescale events
by additionally measuring microlens-parallax values required survey-mode
space-based observation using a satellite in a heliocentric orbit.

\acknowledgments
Work by CH was supported 
by the grants  
of National Research Foundation of Korea (2017R1A4A1015178 and 2019R1A2C2085965).
Work by AG was supported by US NSF grant AST-1516842 and by JPL grant 1500811.
AG received support from the European Research
Council under the European Union's Seventh Framework
Programme (FP 7) ERC Grant Agreement n.~[32103].
The OGLE project has received funding from the National Science Centre, Poland, grant
MAESTRO 2014/14/A/ST9/00121 to AU.
This research has made use of the KMTNet system operated by the Korea
Astronomy and Space Science Institute (KASI) and the data were obtained at
three host sites of CTIO in Chile, SAAO in South Africa, and SSO in
Australia.
The MOA project is supported by JSPS KAKENHI Grant Number JSPS24253004,
JSPS26247023, JSPS23340064, JSPS15H00781, JP17H02871, and JP16H06287.
YM acknowledges the support by the grant JP14002006.
DPB, AB, and CR were supported by NASA through grant NASA-80NSSC18K0274. 
The work by CR was supported by an appointment to the NASA Postdoctoral Program at the Goddard 
Space Flight Center, administered by USRA through a contract with NASA. NJR is a Royal Society 
of New Zealand Rutherford Discovery Fellow.
We acknowledge the high-speed internet service (KREONET)
provided by Korea Institute of Science and Technology Information (KISTI).

\end{document}